\documentclass[prb,aps,twocolumn,amsmath,superscriptaddress]{revtex4}

\usepackage{graphicx}
\usepackage{epsfig}
\usepackage{gensymb}
\begin{document}
\title{Ultrafast acoustic phonon scattering in CH$_3$NH$_3$PbI$_3$ revealed by femtosecond four-wave mixing}

\author{Samuel A. March}
\affiliation{Department of Physics and Atmospheric Science,
Dalhousie University, Halifax, Nova Scotia B3H 4R2 Canada}

\author{Drew B. Riley}
\affiliation{Department of Physics and Atmospheric Science,
Dalhousie University, Halifax, Nova Scotia B3H 4R2 Canada}

\author{Charlotte Clegg}
\affiliation{Department of Physics and Atmospheric Science,
Dalhousie University, Halifax, Nova Scotia B3H 4R2 Canada}

\author{Daniel Webber}
\affiliation{Department of Physics and Atmospheric Science,
Dalhousie University, Halifax, Nova Scotia B3H 4R2 Canada}

\author{Ian G. Hill}
\affiliation{Department of Physics and Atmospheric Science,
Dalhousie University, Halifax, Nova Scotia B3H 4R2 Canada}

\author{Zhi-Gang Yu}
\affiliation{Washington State University, Spokane, Washington 99210 United States}

\author{Kimberley C. Hall}
\affiliation{Department of Physics and Atmospheric Science,
Dalhousie University, Halifax, Nova Scotia B3H 4R2 Canada}


\begin{abstract}
Carrier scattering processes are studied in CH$_3$NH$_3$PbI$_3$ using temperature-dependent four-wave mixing experiments.  Our results indicate that scattering by ionized impurities limits the interband dephasing time (T$_2$) below 30~K, with strong electron-phonon scattering dominating at higher temperatures (with a timescale of 125~fs at 100~K).  Our theoretical simulations provide quantitative agreement with the measured carrier scattering rate and show that the rate of acoustic phonon scattering is enhanced by strong spin-orbit coupling, which modifies the band-edge density of states.  The Rashba coefficient extracted from fitting the experimental results ($\gamma_c=2$~eV\AA) is in agreement with calculations of the surface Rashba effect and recent experiments using the photogalvanic effect on thin films. 
\end{abstract}


\maketitle

The hybrid organic-inorganic perovskite semiconductors (HOIPs) have emerged in recent years as low-cost, solution-processable alternatives to traditional inorganic semiconductors offering good optical and transport properties for applications such as solar cells \cite{NREL:web}, optical sources \cite{Deschler:2014,Zhu:2015,Tan:2014}, photodetectors \cite{Dou:2014}, and field-effect transistors \cite{Kagan:1999,Chin:2015}.  An understanding of the scattering processes that govern the charge carrier dynamics in HOIPs is essential to optimize the performance of such devices.  For instance, electron-phonon scattering determines the rate of relaxation of hot carriers.  The time scale of this process must be well-characterized to evaluate the potential for improvement of solar cell efficiencies through hot carrier extraction.   Scattering with phonons and defects determines the mobility of charge carriers, dictating the efficiency of current extraction in a variety of optoelectronic devices.  Insight into carrier scattering processes has been gained in recent years using time-resolved microwave conductivity \cite{Savenije:2014,Oga:2014}, THz spectroscopy \cite{Karakus:2015,Milot:2015,Cooke:2015}, photocurrent \cite{Phuong:2016}, photoluminescence \cite{Guo:2016,Wright:2016} and transient absorption \cite{Flender:2015,Price:2015,Hopper:2018}.  Such techniques only probe the influence of these scattering processes on the average properties of the carrier distribution (e.g. mobility, carrier temperature).  Four-wave mixing (FWM) techniques provide a direct time-domain probe of carrier scattering \cite{ShahBook,Webber:2017,March:2016,March:2017,Richter:2017,Neutzner:2018}.   Since the fastest scattering process leads to dephasing of electron-hole pairs (with dephasing time $T_2$), FWM enables a determination of the nature and strength of the dominant interactions affecting charge carriers (e.g. carrier-carrier, carrier-phonon, and carrier-defect coupling) in quantitative terms through the measured timescale for coherence decay.  

Previous low-temperature four-wave mixing measurements on the prototypical HOIP (CH$_3$NH$_3$PbI$_3$) revealed weak carrier-carrier scattering in the low carrier density regime ($\leq$10$^{16}$cm$^{3}$) \cite{March:2017}, in stark contrast to inorganic III-V and group IV semiconductors in which carrier-carrier scattering is much stronger than all other carrier interactions \cite{ShahBook}.  This surprising result was attributed to disorder tied to the frozen fluctuations of the methylammonium cations \cite{March:2017}, a static version of the dynamic disorder responsible for large polaron formation and associated screening of the Coulomb interaction at higher temperatures \cite{Zhu:2016}.  Here we report temperature-dependent spectrally-resolved four-wave mixing (FWM) studies with the aim to determine the dominant residual scattering processes in the absence of carrier-carrier scattering.  Our experiments indicate that scattering with phonons dominates for temperatures above 30~K. The rate of phonon scattering we observe for low-energy band carriers is surprisingly high (T$_2$~=~125~fs at a temperature of 100~K).  Simulations of the carrier kinetics incorporating scattering with polar optical and acoustic phonons provide quantitative agreement with our experimental results.  These calculations indicate that the rapid acoustic phonon scattering process we observe is due in part to the strong spin-orbit coupling in this material, which enhances the density of states in the vicinity of the band edge.  For temperatures below 30~K, the measured scattering rate was attributed to ionized impurities.  An estimate of the density of ionized impurities of 1.7$\times$10$^{17}$~cm$^{-3}$ was extracted from the dependence of the carrier scattering rate on electron kinetic energy. Our findings provide insight into electron-phonon interactions and the impact of strong spin-orbit coupling on the fundamental carrier scattering processes in the hybrid perovskite family of materials.

Carrier scattering processes were studied in a thin film of CH$_3$NH$_3$PbI$_3$ in the low-temperature orthorhombic phase using FWM in the two pulse self-diffraction geometry.\cite{ShahBook}  The thin films were prepared using a sequential deposition procedure described previously \cite{March:2016,March:2017,Xiao:2014}.  The FWM technique probes the decay of quantum coherence excited on the optical transitions in the sample using a short, coherent laser pulse.  The decay of the macroscopic polarization density on the interband transitions occurs (with a relaxation time $T_2$) as a consequence of carrier scattering processes that interrupt the phase of the oscillating electric dipoles generated by individual electron-hole pairs.  Since the scattering of either carrier type (electron or hole) leads to loss of coherence, this technique provides a sensitive probe of the fastest scattering process involving charge carriers.  In the FWM technique, two laser pulses $\vec{E_1}(t)$ and $\vec{E_2}(t-\tau)$ propagating with wave vectors $\vec{k_1}$ and $\vec{k_2}$ are used to excite the sample.  The decay of the so-called \textit{self-diffraction} signal along $2\vec{k_2}-\vec{k_1}$ versus interpulse delay $\tau$ indicates the time for loss of coherence on the optical transitions in the sample.    A prism compressor was used for dispersion compensation, resulting in a pulse duration at the sample of 55~fs.  The interpulse delay was varied at 12~Hz using a light-weight retroreflector mounted on a speaker. The four-wave mixing signal along $2\vec{k_2}-\vec{k_1}$ was spectrally-resolved by passing through a 0.25~m monochromator and detected using a photo multiplier tube.  The carrier density determined by the sum of the powers in the two excitation beams, reflection losses and the measured spot size was 2$\times$10$^{16}$ cm$^{-3}$.    

\begin{figure}[htb]\vspace{0pt}
    \includegraphics[width=15.0cm]{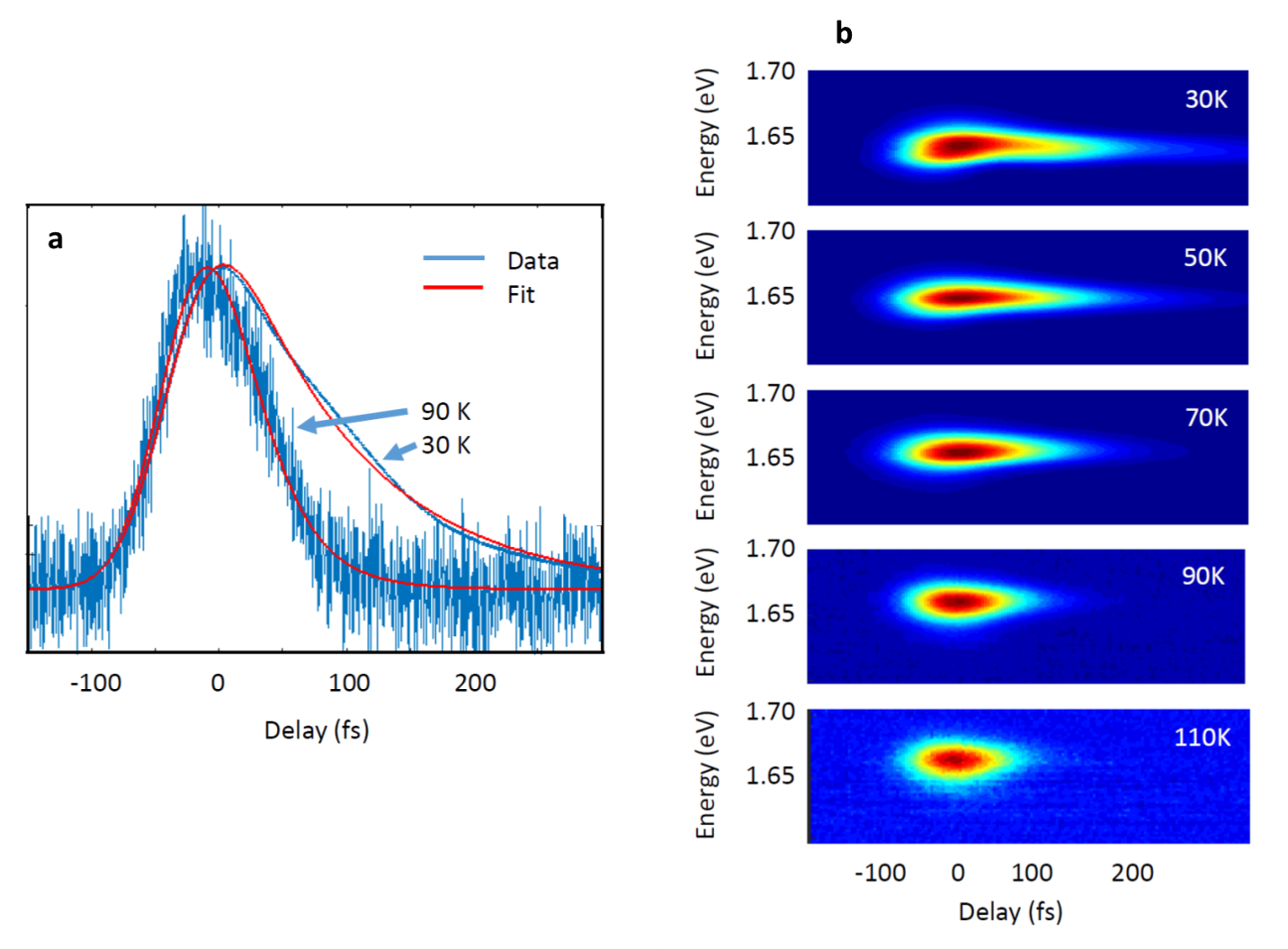}
    \caption{{\bf a} Four-wave mixing signal detected at the band gap of CH$_3$NH$_3$PbI$_3$ for two different temperatures (blue curves), together with fits to a photon echo (red curves), yielding values of $T_2$~=~(400~$\pm$~10)~fs [(125~$\pm$~10~fs)] for 30~K [90~K].   The laser pulse fluence was 1.3 $\mu$J/cm$^{2}$ and the laser was tuned to 1.65~eV. {\bf b} Four-wave mixing signal versus interpulse delay and detection photon energy for various temperatures.  The signal decays more rapidly at higher temperature. } 
    \label{figure1}
\end{figure}

\begin{figure}[htb]\vspace{0pt}
    \includegraphics[width=14.0cm]{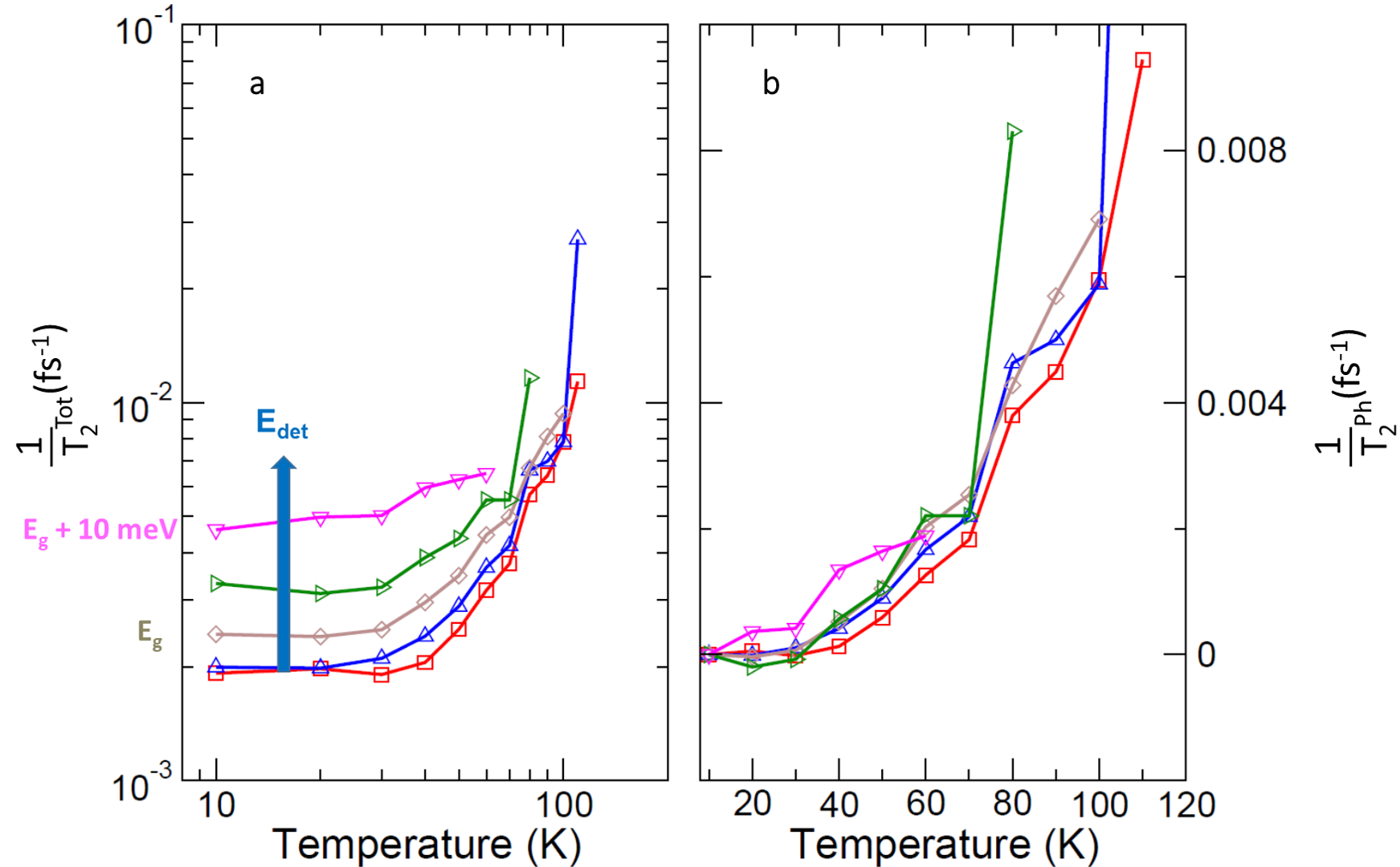}
    \caption{{\bf a} The interband dephasing rate $\frac{1}{T_2}$ obtained from fits to the four-wave mixing signal (symbols) as a function of temperature for detection photon energies ($E_{det}$) ranging from 10 meV below to 10 meV above the band gap at each temperature. {\bf b} Same data as in {\bf a} after subtraction of the dephasing rate at 10~K, attributed to scattering with ionized impurities.  The resulting data indicates the dephasing rate tied to scattering with phonons, which is nearly independent of carrier energy in the vicinity of the band edge but increases rapidly with temperature.}
    \label{figure2}
\end{figure}

The results of FWM experiments on a thin film of CH$_3$NH$_3$PbI$_3$ are shown for two different temperatures in Fig.~\ref{figure1}{\bf a}.  The FWM signal was fit to a photon echo response \cite{YT:1979} convoluted with the laser pulse duration (55~fs) to obtain the interband dephasing time ($T_2$) at each temperature.  The corresponding fits are indicated by the red curves.  The spectrally-resolved FWM signal is shown for a range of temperatures in Fig.~\ref{figure1}{\bf b}.  The weak oscillations in the results for 30~K are tied to excitonic quantum beats, as discussed previously \cite{March:2016}.  These quantum beats disappear and the FWM signal decays more rapidly as the temperature increases, reflecting faster carrier scattering at higher temperatures.  The FWM signal decay time was below the resolution of our experiments for temperatures above 110~K.  The dephasing rate ($\frac{1}{T_2}$) extracted from fits to the FWM signal is shown in Fig.~\ref{figure2}{\bf a} as a function of temperature and detection photon energy. $\frac{1}{T_2}$ increases with increasing carrier energy relative to the band gap, depends weakly on temperature below 30~K and increases rapidly for higher temperatures.    

The relatively constant dephasing rate below 30~K in Fig.~\ref{figure2}{\bf a} is attributed to scattering with ionized impurities.  The rate of ionized impurity scattering may be obtained from the analysis of Conwell and Weisskopf \cite{Conwell:1950} modified to describe interband dephasing rather than momentum relaxation.  (The latter contains a directional factor $1-\cos\theta$ with $\theta$ being the scattering angle; i.e., a forward scattering does not impact the momentum relaxation time, but does reduce $T_2$.)  The resulting dephasing rate may be expressed as $\tau_I^{-1}= \frac{\pi}{4}N^{1/3}_I v$, where $N_I$ is the density of ionzied impurities and $v = \sqrt{\frac{2 E_k}{m}}$ is the carrier velocity and $E_k$ the kinetic energy.  The rate of ionized impurity scattering is independent of temperature and increases with $E_k$.  Using the measured dependence of the dephasing rate at 10~K on the carrier excess energy, we estimate the density of ionized impurities to be 1.7$\times$10$^{17}$cm$^{-3}$. 

If we subtract the 10~K dephasing rate from the measured results at each value of $E_k$, the residual dephasing rate (Fig.~\ref{figure2}{\bf b}) accounts for all other carrier scattering processes. Given the strong increase in the dephasing rate with temperature above 30~K, and the low rate of carrier-carrier scattering in CH$_3$NH$_3$PbI$_3$ at the excitation density used in these experiments (2$\times$10$^{16}$ cm$^{-3}$) \cite{March:2017}, the residual dephasing process is attributed to carrier scattering with phonons.  The overall rate of phonon scattering is faster than that expected for a simple parabolic band given known parameters for the phonon coupling strength and the value of the optical phonon energies \cite{Yu:2017,Yu:2016}.  We show below that both the ultrafast scattering rate and the constant rate versus carrier excess energy are well described by acoustic phonon scattering taking into account the impact of the Rashba effect on the band edge density of states.

The Rashba effect results from the breaking of inversion symmetry in conjunction with spin-orbit coupling \cite{Bychkov:1984,Dresselhaus:1955}, which is large in the hybrid perovskites owing to the incorporation of heavy elements (Pb, I, Br).  While inversion asymmetry is expected in 2D perovskites due to distortions of the lead-iodide octahedra induced by the competing effects of the small and long organic cations \cite{Stoumpos:2016,Todd:2018,Zhai:2017}, 3D perovskites such as CH$_3$NH$_3$PbI$_3$ are believed to possess inversion symmetry in the bulk of the crystal.  Nevertheless, a strong surface Rashba effect is expected in the 3D case  \cite{Mosconi:2017,Che:2018,Frohna:2018,Sarritzu:2018}.  This surface Rashba effect is caused by surface reconstruction that leads to a redistribution of the conduction and valence states resulting in a strong surface dipole \cite{Che:2018} with a penetration depth comparable to typical thin film thicknesses (200-300 nm) \cite{Frohna:2018}.  Recent observations of strong Rashba coupling in single crystal CH$_3$NH$_3$Br$_3$ using the surface-sensitive angle-resolved photoemission spectroscopy technique \cite{Niesner:2016} and in thin films and nanostructures of CH$_3$NH$_3$PbI$_3$ \cite{Isarov:2017,Wang:2017,Rivett:2018,Niesner:2018} are likely dominated by surface Rashba effects.  Such a surface Rashba effect would enable the development of a wide variety of spintronic devices such as spin FETs, all-optical spin switches and spin lasers that would exploit thin film geometries rather than large single crystals \cite{Datta:1990,Hall:2003,Hall:2006,Hall:1999}.  

\begin{figure}[htb]\vspace{0pt}
    \includegraphics[width=15.0cm]{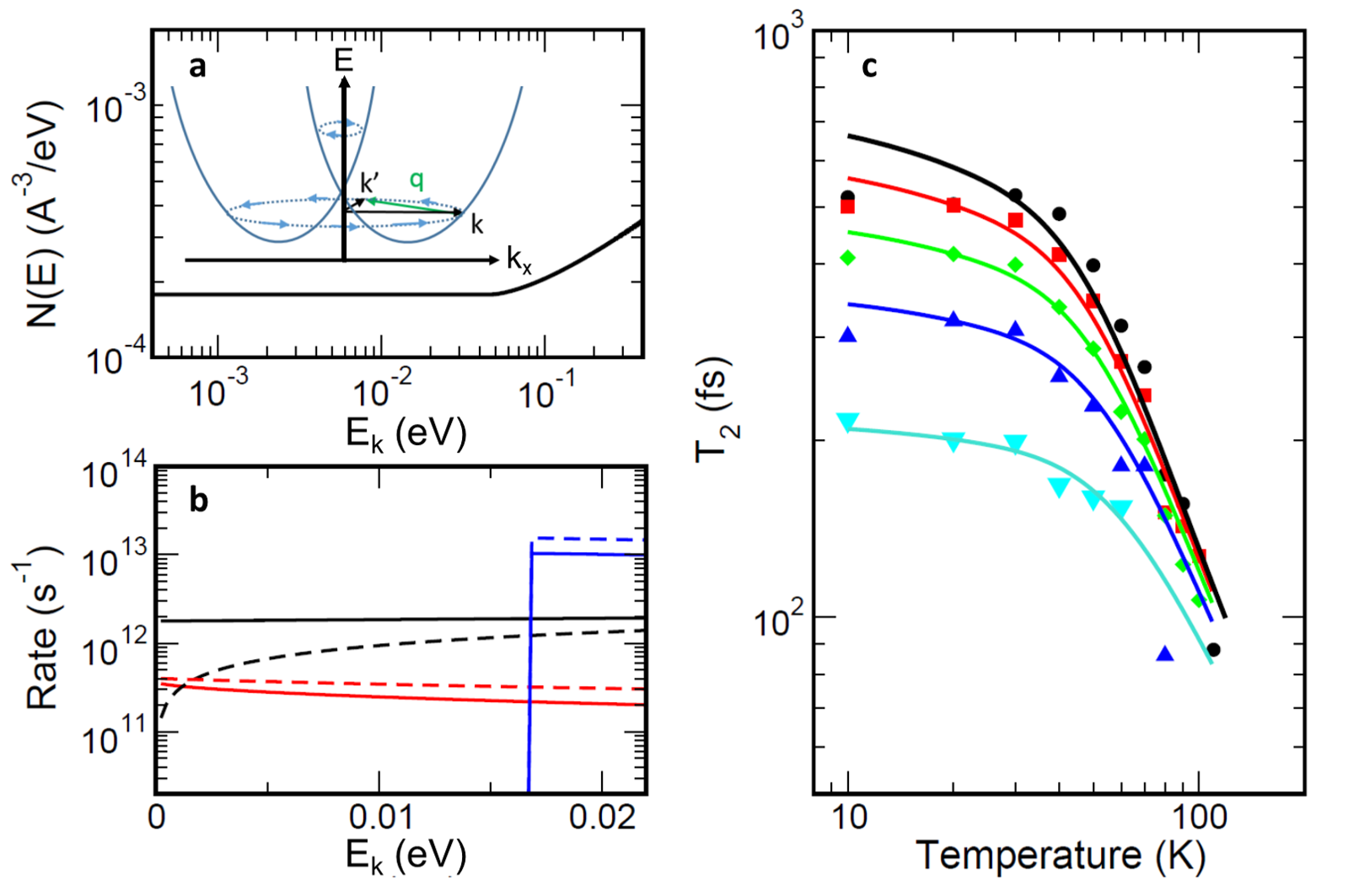}
    \caption{{\bf a}  Calculated density of states in the conduction band including the Rashba splitting.  The DOS is constant for electron energies below the onset of states involving the higher energy spin band, in contrast to the $\sqrt{E}$ dependence for a parabolic band without Rashba coupling. Inset: Schematic electron dispersion relation in the presence of Rashba coupling.  Light blue arrows indicate the direction of the Rashba effective magnetic field as a function of in-plane wavevector, representing the equilibrium spin texture for each spin band.  An elastic scattering event with an acoustic phonon of wavevector $\vec{q}$ (green arrow) in the lower spin band is indicated schematically.  {\bf b} Calculated scattering rate at 50 K as a function of electron energy tied to acoustic phonon scattering (black), and polar optical phonon scattering (red: phonon absorption; blue: phonon emission).  Solid (dashed) curves show the results of calculations with (without) Rashba coupling, showing the strong enhancement of acoustic phonon scattering caused by the Rashba effect.  {\bf c} Results of theoretical simulations of the interband dephasing times (solid curves). The experimental data (symbols) are the same as in Fig.~2{\bf a}.}
    \label{figure3}
\end{figure}

The Rashba effect lifts the degeneracy of the spin states and provides a means to manipulate carrier spin without the need for external magnetic fields \cite{Datta:1990,Hallspintrans:2003,Hall:2006,Hall:1999,Hall:2003}.  The energy-momentum dispersion in the presence of Rashba coupling is:
\begin{equation}
E^{\pm}_{\bm k}=\frac{\hbar^2}{2m_e}[(k_{\perp}\pm k_0)^2+k^2_z], 
\end{equation}
where $m_e$ is the  effective mass, $k_{\perp}=\sqrt{k^2_x+k^2_y}$, $k_0= m \gamma_c /\hbar^2$, $\gamma_c$ is the Rashba coupling strength, and $E^0_{\rm c}-\frac{\hbar^2 k_0^2}{2m_e} \equiv 0$. The corresponding eigenstates are $|\psi_{\pm}({\bm k}) \rangle =\frac{1}{\sqrt{2}}(1,\pm ie^{i\phi})^T$ with the spin orientation being a function of the momentum direction.  A schematic representation of the spin-split states in the conduction band is shown in the inset to Fig.~\ref{figure3}{\bf a}.   The Rashba effect shifts the band edges away from the high symmetry point in the corresponding inversion symmetric structure, leading to an indirect band gap separated from the direct gap by a relatively small energy ($\sim$50 meV) \cite{Wang:2017,Hutter:2016}.  This feature has been invoked to explain the simultaneous presence of a large absorption coefficient and long carrier recombination time in CH$_3$NH$_3$PbI$_3$ thin films leading to high solar cell efficiencies \cite{Amat:2014,Motta:2015,Etienne:2016,Mosconi:2017,Zheng:2015,Kirchartz:2017,Yu:2017}.  Much less is known about the effect of Rashba coupling on the scattering and relaxation processes in HOIPs \cite{Yu:2017,Even:Review,Yu:2016}, despite the crucial role played by these scattering processes in the operation of optoelectronic devices.  

In order to explain the measured temperature dependence of $T_2$, the dephasing rate tied to scattering of electrons with acoustic and polar optical phonons was calculated following a model developed previously \cite{Yu:2016} taking into account the Rashba splitting.  The electron-phonon coupling can be expressed as  
\begin{equation}
H_{\rm ep}=\sum_{{\bm kq}s}(V_{\bm q}b_{\bm q}c^{\dag}_{{\bm k}+{\bm q}s}c_{{\bm k}s}+V^*_{\bm q}b^{\dag}_{\bm q}c^{\dag}_{{\bm k}s}c_{{\bm k}-{\bm q}s})
\end{equation}
where $c^{\dag}_{{\bm k}s}$ creates an electron with momentum ${\bm k}$ and spin $s$ $(=\uparrow,\downarrow)$, $b^{\dag}_{\bm q}$ creates a phonon with momentum ${\bm q}$, and the two terms describe phonon absorption and emission processes. Since $H_{\rm ep}$  does not change electron spin,
the matrix elements between electron eigenstates,  $M_{\pm\pm}({\bm k},{\bm k}')\equiv \langle \psi_{\pm}({\bm k}')|H_{\rm ep}|\psi_{\pm}({\bm k})\rangle$,  in the presence of  the Rashba effect are 
\begin{eqnarray}
M_{++}({\bm k},{\bm k}')&=&M_{--}({\bm k},{\bm k}')=V_{{\bm k}'-{\bm k}}\frac{1+e^{i(\phi'-\phi)}}{2},\\
M_{+-}({\bm k},{\bm k}')&=&M_{-+}({\bm k},{\bm k}')=V_{{\bm k}'-{\bm k}}\frac{1-e^{i(\phi'-\phi)}}{2}.
\end{eqnarray}   The coupling between electrons and acoustic phonons is 
\begin{equation}
V_{\bm q}=i E_1 \Big (\frac{\hbar}{2 \rho \omega_{\bm q}}\Big)^{1/2} q \equiv i E_1 \Big (\frac{\hbar v_s}{2 c_L}\Big)^{1/2}\sqrt{q}
\end{equation}
where $E_1$ is the deformation potential coupling for the conduction band, $\rho$ is the material density, 
$\omega_{\bm q} = v_s q$ is the acoustic phonon dispersion with $v_s$ being the velocity of sound in the material, and $c_L=\rho v^2_s$ is the longitudinal elastic  constant. In HOIPs, the polar coupling for electrons can be expressed as
\begin{equation}
V_{\bm k}=-i\sum_{j}\Big(\frac{4\pi\alpha_{{\rm e}j}}{\Omega} \Big)^{1/2}\frac{\hbar \omega_{l_j}}{q}\Big(\frac{\hbar}{2m_{\rm e}\omega_{l_j}}\Big)^{1/4}
\end{equation}
where $\alpha_{{\rm e}j}$ is the dimensionless polar coupling strength for the jth optical phonon mode \cite{Yu:2016}.  The scattering rate can be calculated from the Fermi's golden rule.  In our calculations, we restrict ourselves to the scattering rate of electrons.  As we show below, the Rashba effect strongly enhances the rate of acoustic phonon scattering.  Since the magnitude of the Rashba effect in the valence band is expected to be weaker than in the conduction band due to the dominant contribution of the I 5p orbital to the valence states\cite{Even:Review}, electron scattering is assumed to dominate the interband dephasing process.  The magnitude of the Rashba coupling strength $\gamma_c$ was taken as an adjustable parameter to obtain the best fit to the experimental results.     

The results of theoretical simulations of the interband dephasing rate are shown in Fig.~\ref{figure3}{\bf c}, together with the experimental data.  For these simulations, only one optical phonon mode is included tied to the Pb-I stretching mode, which was shown previously to have the largest polar optical coupling ($\alpha_{\rm e}=1.1$) with phonon energy $\hbar\omega_0=16.7$ meV \cite{Yu:2016}, and the value of the Rashba coupling strength is $\gamma_c=2$~eV\AA.  The theoretical simulations provide good quantitative agreement with the experimental results, including both the temperature dependence and the lack of dependence of the phonon scattering rate on carrier energy.  The contribution of each phonon scattering process to the calculated dephasing rate at a temperature of 50~K is shown as a function of electron energy in Fig.~\ref{figure3}{\bf b}.  The solid (dashed) curves indicate the scattering rates with (without) Rashba coupling.  Acoustic phonon scattering dominates over polar optical phonon scattering for the full range of temperatures considered here ($\leq$110~K).    

The rate of acoustic phonon scattering is strongly enhanced for electrons energies near the band edge with Rashba coupling included. This strong enhancement is the origin of the ultrafast interband dephasing observed in our experiments.  Our calculations indicate an enhancement of the low temperature acoustic phonon scattering rate of 6$\times$ relative to the case of a simple parabolic band without Rashba coupling.  The Rashba effect also modifies the energy-dependence of the scattering rate tied to acoustic phonons, which changes from an approximately square-root dependence without Rashba to constant with the Rashba effect included.  This feature is tied to the constant density of states in the lower-energy spin band (Fig.~\ref{figure3}{\bf a}), and explains the lack of an observed dependence of the electron-phonon scattering contribution to the dephasing rate on carrier energy near the band gap. The $\sqrt{q}$ dependence of the acoustic phonon coupling strength also contributes to the enhancement of the scattering rate as the Rashba effect leads to an increase in the average $|q|$ for phonon scattering near the band minimum (Fig.~\ref{figure3}{\bf a}, inset).  The value of $\gamma_c=2$~eV\AA we extract from our simulations is in good agreement with a recent calculation of the surface Rashba effect using density functional theory \cite{Mosconi:2017} as well as recent observations in CH$_3$NH$_3$PbI$_3$ films in the orthorhomic phase using the circular photogalvanic effect \cite{Niesner:2018}.

In conclusion, we have applied four-wave mixing to measure electron scattering times in thin films of CH$_3$NH$_3$PbI$_3$.  Our experiments indicate that ionized impurities provide the dominant scattering process at low temperatures, while above 30~K strong temperature-dependent acoustic phonon scattering dominates the carrier kinetics.  Our experiments reveal that the rate of scattering with phonons is quite high, even for electrons very close to the band edge, and that the electron-phonon scattering rate is relatively insensitive to the carrier energy.  Both of these findings were accounted for in numerical simulations taking into account the influence of the Rashba effect on the density of states near the band edge.  These simulations yield a Rashba coupling strength of $\gamma_c=2$ eV\AA, believed to be dominated by inversion symmetry breaking at surfaces and interfaces in the polycrystalline thin film.  The insight into the strength of electron-phonon coupling provided by FWM experiments in this work complements recent broadband transient absorption experiments revealing phonon-induced modulation of the carrier transition energies \cite{Ghosh:2017}.  Our findings provide insight into carrier scattering processes in HOIPs with implications for a wide range of device applications.
  
\clearpage
\clearpage
\clearpage
\clearpage


This research is supported by the Natural Sciences and Engineering Research Council
of Canada.  Work at Washington State University was supported by grant W911NF-17-1-0511 from the US Army Research Office.

\end{document}